\documentclass[screen,sigconf]{acmart}
\usepackage{graphicx}
\usepackage{xspace}
\usepackage{listings}
\usepackage{extdash}
\usepackage{textcomp}
\usepackage{fancyhdr}
\usepackage{booktabs}
\usepackage{longtable}
\usepackage{url}
\usepackage{enumitem}
\usepackage{lipsum}

\usepackage[]{siunitx}
\DeclareSIUnit\flop{FLOP}
\DeclareSIUnit\flops{FLOPs}
\DeclareSIUnit\billion{B}
\DeclareSIUnit\million{M}
\usepackage{bm}

\definecolor{RYB1}{RGB}{207, 37, 37}
\definecolor{RYB2}{RGB}{37, 91, 207}
\definecolor{RYB3}{RGB}{37, 207, 91}
\definecolor{RYB4}{RGB}{163,26,145}
\definecolor{RYB5}{RGB}{253, 180, 98}
\definecolor{RYB6}{RGB}{179, 222, 105}
\definecolor{RYB7}{RGB}{128, 177, 211}

\definecolor{lightblue}{rgb}{0.63, 0.74, 0.78}
\definecolor{seagreen}{rgb}{0.18, 0.42, 0.41}
\definecolor{orange}{rgb}{0.85, 0.55, 0.13}
\definecolor{silver}{rgb}{0.69, 0.67, 0.66}
\definecolor{rust}{rgb}{0.72, 0.26, 0.06}
\definecolor{purp}{RGB}{68, 14, 156}

\colorlet{lightrust}{rust!50!white}
\colorlet{lightorange}{orange!25!white}
\colorlet{lightlightblue}{lightblue}
\colorlet{lightsilver}{silver!30!white}
\colorlet{darkorange}{orange!75!black}
\colorlet{darksilver}{silver!65!black}
\colorlet{darklightblue}{lightblue!65!black}
\colorlet{darkrust}{rust!85!black}
\colorlet{darkseagreen}{seagreen!85!black}

\usepackage[nameinlink]{cleveref}
\crefname{lstlisting}{listing}{listings}
\Crefname{lstlisting}{Listing}{Listings}

\setcopyright{acmlicensed}
\copyrightyear{2025}
\acmYear{2025}
\setcopyright{cc}
\setcctype{by}
\acmConference[SC Workshops '25]{Workshops of the International Conference for High Performance Computing, Networking, Storage and Analysis}{November 16--21, 2025}{St Louis, MO, USA}
\acmBooktitle{Workshops of the International Conference for High Performance Computing, Networking, Storage and Analysis (SC Workshops '25), November 16--21, 2025, St Louis, MO, USA}
\acmDOI{10.1145/3731599.3767424}
\acmISBN{979-8-4007-1871-7/2025/11}

\begin{document}

\citestyle{acmnumeric}

\title[Testing and benchmarking emerging supercomputers via the MFC flow solver]{
    Testing and benchmarking emerging supercomputers via the MFC flow solver
}

\author{Benjamin Wilfong}
\affiliation{%
  \institution{Georgia Institute of Technology}
  \city{Atlanta}
  \state{Georgia}
  \country{USA}
}

\author{Anand Radhakrishnan}
\affiliation{%
  \institution{Georgia Institute of Technology}
  \city{Atlanta}
  \state{Georgia}
  \country{USA}
}

\author{Henry A.\ Le Berre}
\affiliation{%
  \institution{Georgia Institute of Technology}
  \city{Atlanta}
  \state{Georgia}
  \country{USA}
}

\author{Tanush Prathi}
\affiliation{%
  \institution{Georgia Institute of Technology}
  \city{Atlanta}
  \state{Georgia}
  \country{USA}
}

\author{Stephen Abbott}
\affiliation{%
  \institution{Hewlett Packard Enterprise}
  \city{Bloomington}
  \state{Minnesota}
  \country{USA}
}

\author{Spencer H.\ Bryngelson}
\affiliation{%
  \institution{Georgia Institute of Technology}
  \city{Atlanta}
  \state{Georgia}
  \country{USA}}
\authornotemark[1]
\email{shb@gatech.edu}

\renewcommand{\shortauthors}{Wilfong et al.}

\begin{abstract}
    Deploying new supercomputers requires testing and evaluation via application codes.
    Portable, user-friendly tools enable evaluation, and the Multicomponent Flow Code (MFC), a computational fluid dynamics (CFD) code, addresses this need.
    MFC is adorned with a toolchain that automates input generation, compilation, batch job submission, regression testing, and benchmarking.
    The toolchain design enables users to evaluate compiler--hardware combinations for correctness and performance with limited software engineering experience.
    As with other PDE solvers, wall time per spatially discretized grid point serves as a figure of merit.
    We present MFC benchmarking results for five generations of NVIDIA GPUs, three generations of AMD GPUs, and various CPU architectures, utilizing Intel, Cray, NVIDIA, AMD, and GNU compilers.
    These tests have revealed compiler bugs and regressions on recent machines such as Frontier and El~Capitan.
    MFC has benchmarked approximately 50 compute devices and 5 flagship supercomputers.
\end{abstract}



\maketitle

\section{Introduction}

Supercomputer tests and benchmarks rely on portable and performant software applications to make meaningful comparisons between new and existing systems and hardware.
This work presents MFC~\cite{bryngelson21} as an application that addresses this need.
MFC is a GPU-accelerated~\cite{radhakrishnan24}, feature-rich~\cite{wilfong252}, portable~\cite{wilfongSC24}, and user-friendly computational fluid dynamics (CFD) code used to test and benchmark five generations of NVIDIA GPUs, three generations of AMD GPUs, and many CPUs.

MFC is a multiphysics flow solver that has been used to simulate compressible multi-species, phase, and chemically reacting fluid flows~\cite{charalampopoulos21,bryngelson19_whales,bryngelson23,cisneros25,panchal23}.
The spatiotemporal requirements of compressible multiphysics flow simulations have driven the authors to prioritize performance and portability in the design of MFC.
The background of researchers in the field has also led us to prioritize user-friendliness and approachability in MFC's design.
This work describes how MFC's predictable performance and user-friendly interface make it a reliable and approachable application for testing and benchmarking new supercomputers.

The MFC toolchain is designed to be user-friendly and portable, helping users test and benchmark HPC systems with minimal knowledge of the underlying hardware or software.
The user interacts with the toolchain via a wrapper script that requires a one-time setup to add support for an alternative system.
The user completes setup by identifying and specifying the required Lmod~\citep{mclay2011lmod} modules and shell environment variables.
The final step is to create a system-specific job-submission template file that supports multiple schedulers and their idiosyncrasies.
The bash wrapper automates the process of loading modules, building MFC and its dependencies, running regression tests, and benchmarking the code once the initial setup is complete.
The toolchain is designed to be easily adapted and updated by users.
This strategy lets users modify test cases and benchmarks to evaluate code and language features, confirm system-specific correctness, and identify performance bottlenecks.

\begin{figure*}[htpb]
    \Description{A block diagram showing the connectivity of the MFC toolchain.}
    \centering
    \includegraphics[]{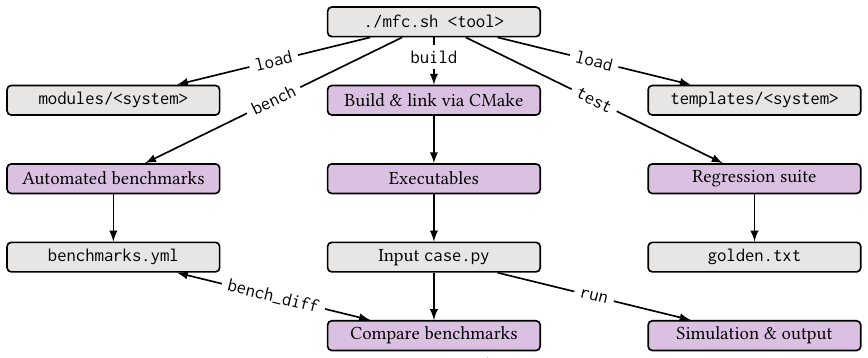}
    \caption{The MFC toolchain and its connectivity.}
    \label{fig:overview}
\end{figure*}

We summarize MFC's performance with a single figure of merit, \textit{grindtime}: nanoseconds of wall time per grid point, equation, and right-hand-side evaluation.
Here, the grid points represent the spatial discretization points of the simulation domain, the equations refer to the system of partial differential equations solved by the code, and the right-hand side evaluations denote the operations performed to advance the solution in time.
This definition provides a figure that describes the time it takes to perform the smallest measurable unit of work in an application that solves time-dependent partial differential equations (PDEs).
Defining the grindtime in terms of wall time means it follows strong scaling trends when increasing device count.
We compare a single GPU driven by one MPI rank with a single CPU die driven by multiple.
Expressing performance per smallest measurable unit makes grindtime independent of problem size, the number of physical model equations, and the time-integration scheme.

The grindtime measurement accounts for MPI communication and host--device transfers relevant to network, CPU, and offload device (e.g., GPU) performance.
It neglects the time spent performing code initialization and I/O operations.
I/O costs are not directly benchmarked in the present work as they are sufficiently small compared to compute costs.
Still, MFC writes an I/O profile for each case, which can be used to evaluate I/O performance or bottlenecks if unexpected behavior is observed.
Defining grindtime in this way evaluates how hardware and network performance impact the run-time of core compute kernels.

\subsubsection*{Manuscript structure:}

MFC's suggested role in the existing landscape of HPC testing and benchmarking tools is described in \cref{sec:existingTools}.
\Cref{sec:summary} describes the steps for testing and benchmarking a supercomputer with MFC.
\Cref{sec:correctness} and \cref{sec:performance} provide details on how users can extend the testing and benchmarking capabilities of MFC and give examples of bugs and performance bottlenecks identified by each tool.
Application of MFC as a tool for system deployment and testing via a standardized benchmark case, weak scaling, and strong scaling is described in \cref{sec:exampleUses}.
Limitations, implications, and final thoughts are given in \cref{sec:limitations} and \cref{sec:conclusion}.

\section{Existing tools} \label{sec:existingTools}

Many tools support the benchmarking of HPC systems.
One such tool is the High-Performance Linpack (HPL) benchmark~\cite{dongarra2003linpack}, which solves a dense system of linear equations via LU factorization with partial pivoting.
The HPL benchmark estimates a supercomputer’s maximum sustained performance, which informs the semiannual TOP500 ranking~\cite{dongarra2003linpack}.
HPL is widely accepted but limited in scope, representing only a fraction of user applications.
The SPEChpc benchmark suite~\cite{specHPC21} was created, in part, as a response to this limitation.
The SPEChpc benchmarks comprise multiple maintained user applications that vary in subject matter, numerical methods, and programming models, providing a more holistic measure of system performance.

Benchmarking and automation tools can help users test HPC systems.
A non-exhaustive list of such tools includes ReFrame~\cite{reframe}, JUBE~\cite{luhrs2016jube}, Ramble~\cite{ramble}, BenchPRO~\cite{benchpro}, and OLCF~Test Harness~\cite{olcfTestHarness}.
Each tool has strengths, but all aim to provide automated interfaces for testing and benchmarking supercomputers and their hardware.
MFC itself is not a suitable replacement for any of these tools.
MFC is, however, a suitable application for testing and benchmarking HPC systems with these tools.
The user-friendly interface and automated building, testing, and benchmarking processes in MFC make it an easy-to-integrate candidate for use with these existing tools.

\section{Tooling summary and usage}\label{sec:summary}

The core of MFC's user-friendly interface is the bash wrapper \texttt{mfc.sh} that provides quick access to all of MFC's automated tools.
The control flow of the toolchain is shown in \cref{fig:overview}, and \cref{tab:clTools} lists the tools for testing and benchmarking a new system.
The following sections describe how a user would typically use each of these tools to test and benchmark a new system using MFC.

\begin{table}
    \centering
    \caption{List of relevant automated tools accessible via the wrapper script \texttt{mfc.sh}.
    The tools in this list are in the order a user would typically use them to test and benchmark a new system with MFC.
    Each tool has additional command-line options accessible via \texttt{./mfc.sh <tool> -{}-help}.}
    \label{tab:clTools}
    \begin{tabular}{l l}
        \toprule
        Tool & Description \\ \midrule
        \texttt{load} & Load modules and initialize environment \\
        \texttt{build} & Build MFC's source and dependencies \\
        \texttt{test} & Run the regression test suite \\
        \texttt{bench} & Run the benchmark suite \\
        \texttt{bench\_diff} & Compare benchmark results \\
        \texttt{run} & Run a user-defined case file \\
        \bottomrule
    \end{tabular}
\end{table}

\subsubsection*{Step 1: System and environment setup.}

The first step in testing and benchmarking a new system with MFC is to set up the compute environment.
Setting up the environment starts with identifying the relevant modules to load and environment variables to set.
The required modules for currently supported clusters and supercomputers are listed in \texttt{toolchain/modules} and are easy to extend.
\Cref{lst:deltaModules} shows an example entry for NCSA~Delta, which supports CPU and GPU builds with different modules and environment variables.
Line~1 assigns the system name to the identifier \texttt{d} to be used in the command line interface.
Modules and environment variables used by both CPU and GPU builds are stored in the \texttt{d-all} entry and loaded first.
The \texttt{d-cpu} and \texttt{d-gpu} entries store modules and environment variables specific to building and running MFC on CPU or GPU hardware.
Once the relevant modules and environment variables are identified, the user can load them with the command \texttt{source ./mfc.sh load}.
This command prompts the user for the system identifier (for example \texttt{(d)} for NCSA~Delta), and configuration (\texttt{(c | cpu)} for CPU builds and \texttt{(g | gpu)} for GPU builds).
Executing \texttt{source ./mfc.sh load} is configured to purge loaded modules and load the modules and environment variables appropriate for that system.

\begin{lstlisting}[
    language=bash,
    caption={
        Example module and environment variables loaded into the user's environment for NCSA~Delta.
    },
    label={lst:deltaModules},
    xleftmargin=2em,
    numbers=left]
d      NCSA Delta
d-all python/3.11.6
d-cpu gcc/11.4.0 openmpi
d-gpu nvhpc/24.1 cuda/12.3.0 openmpi/4.1.5+cuda cmake
d-gpu CC=nvc CXX=nvc++ FC=nvfortran
d-gpu MFC_CUDA_CC=80,86
\end{lstlisting}

The final step in setting up the environment is to create a system-specific template file in the \texttt{toolchain/templates/} directory.
MFC uses the \texttt{Mako} templating library~\cite{mako} for system-specific templates that create Bash scripts, which can run MFC executables in interactive or batch mode, depending on the user's needs.

System-specific templates are used because they provide a way to support multiple scheduling systems, such as Slurm, PBS, LSF, and Flux, without requiring future users to be familiar with the details of the scheduling system.
Templates can also set additional run-time environment variables and settings that are irrelevant to compilation.
For example, the template for OLCF~Frontier sets the environment variable \texttt{MPICH\_GPU\_SUPPORT\_ENABLED=1} to activate GPU-aware MPI at run-time.
The template also sets \texttt{ulimit -s unlimited} to expand the stack size for simulating particularly large problems.
The template files help the toolchain add the commands to perform high-level and kernel-level profiles to interactive and batch scripts.
This approach removes the need for direct user interaction with profiling tools.
Once the module and template files are created for the new system, one can build MFC and its dependencies.

\subsubsection*{Step 2: Building.}

Once the relevant modules are loaded and environment variables set, the user can build MFC and its dependencies with the command \texttt{./mfc.sh build <config>}, where \texttt{<config>} is \texttt{-{}-gpu acc|mp} for GPU builds and \texttt{-{}-no-gpu} for CPU builds.
GPU acceleration is provided via OpenACC~\cite{OpenACC}, which supports GPU offloading for NVIDIA and AMD GPUs, or OpenMP~\cite{OpenMP}, the latter being better supported by compilers targeting AMD and Intel GPUs.
MFC's dependencies vary by system and hardware.
All MFC simulations depend on \texttt{silo} and \texttt{hdf5} for visualization and post-processing.
\texttt{silo} and \texttt{hdf5} are fetched and compiled automatically by CMake for the specific hardware and system configuration.
Some features in MFC rely on fast Fourier transforms (FFTs), which are provided by \texttt{FFTW}~\cite{FFTW} for CPU builds, \texttt{cuFFT}~\cite{cuFFT} for NVIDIA GPU builds, and \texttt{hipFFT}~\cite{hipFFT} for AMD GPU builds.
CMake automatically detects the hardware and system configuration and builds the appropriate Fourier transform library.
Then, the user tests the build.

\subsubsection*{Step 3: Regression testing:}

With the system environment set up and the source code built, MFC's test suite can now be run to identify any run-time errors or correctness bugs resulting from the hardware-software configuration.
The test suite is run with the command \texttt{./mfc.sh test -{}- -c <system>} where \texttt{<system>} is the name of the \texttt{mako} template file created in step 1.
At the time of writing, MFC's test suite comprises approximately 500 test cases that cover most available features.
\Cref{sec:correctness} describes details of the regression suite, including how cases are added, how results are compared, and how the suite has aided in identifying and correcting compiler-hardware bugs.
The user can proceed to the benchmarking step once a hardware-compiler combination has been tested for correctness.

\subsubsection*{Step 4: Benchmarking.}

MFC contains two benchmarking tools that can be used to measure the performance of hardware-compiler combinations.
The first is a standardized benchmark case with documented performance on 49 different hardware platforms, providing a quick way to compare performance across systems.
This benchmark case is described in more detail in~\cref{sec:standardBenchmark}.
The second benchmarking tool is an automated benchmark suite that tests a broader range of MFC's functionality and summarizes the performance for each test in a single \texttt{yaml} file.
The benchmarks are executed via \texttt{./mfc.sh bench -{}-mem <gb/rank> -o <output>.yml -{}- -c <system> -n <nranks> <device\_opts>}.
The command line arguments are described in \cref{tab:benchArgs}.

\begin{table}[h]
    \centering
    \caption{Arguments for the automated benchmarks.}
    \begin{tabular}{r l p{1.9in}}
        \toprule
        Flag & Argument & Description \\
        \midrule
        \texttt{-{}-mem} & \texttt{<gb/rank>} & Memory (in GB) of problem size per rank \\
        \texttt{-o} & \texttt{<output>} & Output \texttt{yaml} file with summary results \\
        \texttt{-c} & \texttt{<system>} & Name of \texttt{Mako} template (Step 1) \\
        \texttt{-n} & \texttt{<nranks>} & MPI ranks used for benchmarking \\
        n/a & \texttt{<device\_opts>} & \texttt{-{}-gpu} (GPU); \texttt{-{}-no-gpu} (CPU) \\
        \bottomrule
    \end{tabular}
    \label{tab:benchArgs}
\end{table}

The performance results for the benchmarks on different systems can be compared and summarized by running \texttt{./mfc.sh bench\_diff <ref\_ouput>.yml <output>.yml}, where \texttt{<ref\_output>.yml} is the output file from running the benchmark suite on a reference system.
After steps 1--4, the user can be confident that the hardware--compiler combination produces correct results and expected performance, and proceed to use MFC for their intended scientific use case.

\subsubsection*{Step 5: Running.}

With correctness and performance verified, the user can create a user-defined case file and run it with \texttt{./mfc.sh run}.
The details for creating and running user-defined cases are beyond the scope of this work, but are described in detail in the MFC documentation\footnote{\url{mflowcode.github.io/documentation}}.

\section{Regression test details} \label{sec:correctness}

The MFC regression suite tests over 500 unique cases (at the time of writing) and is designed to be readily extended and maintained by users.
Each test case is based on a generic case file that can be modified using a stack mechanism to add or replace variables, enabling or disabling any MFC feature.
Once a case is defined, an eight-digit universally unique identifier (UUID) is associated with it, and a directory in the source is created to store the case's golden file and associated metadata, including device information and build configuration, directly via CMake.
The user can create the golden file and metadata by running the test with \texttt{./mfc.sh test -o <UUID> -{}-generate}, where \texttt{<UUID>} is the unique identifier for the test case.
More information on each of these steps is provided below.

\subsection{Test case definition}

\Cref{lst:testCase} demonstrates how a new codebase feature can be added to the MFC test suite.
In this case, we call \texttt{alter\_igr()} to create a span of tests that cover the relevant adorning features of MFC; 24 are made here by calling \texttt{alter\_igr()} with six unique base stacks.
Line~\ref{line:2} adds \texttt{`igr': `T'} and three other parameters to run all test cases that use it to the case stack, and adds `IGR' to the human-readable trace that describes the case.
Line~\ref{line:4} defines a loop over the available numerics added to the case stack and human-readable trace in line~\ref{line:5}.
Lines~\ref{line:6} and~\ref{line:7} define the test cases for two available iterative linear solvers via the function \texttt{define\_case\_d()} with the case stack, a human-readable trace entry, and a dictionary of additional variables to set in the case file.
The contents of lines~\ref{line:2} and~\ref{line:5} are popped from the stack in lines~\ref{line:9} and~\ref{line:11} and return the stack to its original state.
The stack-based approach enables ready extension and modification of the test suite without requiring user knowledge of other code features or their functionality.
The human-readable trace appended by the function \texttt{alter\_igr()} is printed to the command line, along with the case UUID, allowing the user to identify the case and its parameters.

\begin{lstlisting}[
    language=python,
    caption={
        Code snippet demonstrating how the stack-based approach adds a new code feature to the test suite.
    },
    label={lst:testCase},
    xleftmargin=2em,
    numbers=left,
    float=*,
    floatplacement=tbp,
    escapeinside={(*@}{@*)}]
def alter_igr():(*@\label{line:1}@*)
    stack.push('IGR',{'igr': 'T',  'alf_factor': 10, 'num_igr_iters': 10, 'num_igr_warm_start_iters': 10})(*@\label{line:2}@*)
(*@\label{line:3}@*)
    for order in [3, 5]:(*@\label{line:4}@*)
        stack.push(f"igr_order={order}", {'igr_order': order})(*@\label{line:5}@*)
        cases.append(define_case_d(stack, 'Jacobi', {'igr_iter_solver': 1}))(*@\label{line:6}@*)
        if order == 5:(*@\label{line:7}@*)
            cases.append(define_case_d(stack, 'Gauss Seidel', {'igr_iter_solver': 2}))(*@\label{line:8}@*)
        stack.pop()(*@\label{line:9}@*)
(*@\label{line:10}@*)
    stack.pop()(*@\label{line:11}@*)
\end{lstlisting}

\subsection{Golden file generation and maintenance}

Golden files are reference output data used to verify the correctness of numerical simulations by comparing current test results against previously validated solutions.
With MFC's toolchain, they are created, along with their associated metadata, by running the test case as \texttt{./mfc.sh test -o <UUID> -{}-generate}, where \texttt{<UUID>} is the case identifier.
This command executes the test, creating the golden file \texttt{golden.txt} and the metadata file \texttt{golden-metadata.txt}.
These files use MFC's serial output formatting and record the CMake configuration, system information, and hardware information.
Each line in \texttt{golden.txt} contains a flattened array storing a single simulation output.
This format enables easy comparison of the output of different systems while minimizing the size of the golden files in version control.
Once a golden file is created, test suite execution compares the code output to the golden file.
It reports instances where an absolute or relative error exceeds a user-defined threshold.
By default, an absolute and relative test tolerance of $1\times 10^{-12}$ is used for double precision computations.
This tolerance reflects floating-point round-off and non-IEEE-754-compliant optimized floating-point operations at run time.

If test cases are changed so that other outputs must be added to the golden file, the user can update it.
One updates golden files as \texttt{./mfc.sh test -o <UUID> -{}-add-new-variables}.
Executing this command adds new tracked variables to the golden file without modifying the existing values, thereby maintaining the integrity of the original data.

\subsection{System and compiler bugs identified}

The ability to build and test code that leverages a wide breadth of modern Fortran's features has enabled developers to identify and file tickets for at least 15 compiler and system bugs and regressions.
The most common bugs identified are related to improper handling of module variables and unexpected behavior of the \texttt{!\$acc routine seq} and \texttt{host\_data use\_device} directives.
Once a bug is identified, a minimum working example can be created to open formal tickets with compiler vendors, aiding in their timely resolution.
These bugs are relevant to core features of the Fortran language and OpenACC offloading, making their identification valuable to both MFC developers and the broader HPC community.
Automated testing across core language features has highlighted idiosyncrasies among compilers using the same GPU offload model.
The automated test suite has also proven invaluable in identifying bugs in the MFC source code itself, often due to refactoring that improves the overall quality of the source.

\section{Performance test details}\label{sec:performance}

At the time of writing, MFC's automated benchmark suite contains five test cases that cover its most commonly used features.
Test cases for new features can be added easily to the benchmark suite by creating a new directory with a case file in the \texttt{benchmarks/} directory and adding the case to the \texttt{toolchain/benchmarks.yml} file.
Each benchmark case accepts an argument defining the approximate problem size per rank in gigabytes of memory and automatically scales to any number of MPI ranks.
The summary data for each benchmark case is stored in a \texttt{yaml} file, which contains the total wall time (in seconds), grindtime, and a summary of the invocation used to run the benchmark.
The relative performance of two systems can be compared automatically using MFC's \texttt{bench\_diff} tool, which prints a human-readable summary table.
All benchmark cases use MFC's \texttt{-{}-case-optimization} flag to specify certain case parameters as compile-time constants, enabling more aggressive compiler optimizations.
The \texttt{-{}-case-optimization} flag results in approximately a ten-fold improvement in grindtime performance, though speedup varies depending on the compiler and hardware used.

\subsection{Performance bottlenecks identified}

The automated benchmark suite has identified several performance regressions and bottlenecks, particularly on new architectures.
The largest performance impacts occur when compilers cannot inline subroutines called within GPU compute regions.
MFC uses negative indices to simplify array indexing when needed.
When allocatable variables with non-default lower bounds are not defined at compile time, NVIDIA's NVHPC compiler does not mark them for inlining unless the compiler flag \texttt{-Minline=reshape} is specified, enabling speedups.
Such subroutines are easily introduced when attempting to refactor code, so automatically identifying the resulting performance regression is valuable in maintaining MFC's performance.
We observe additional inlining-related performance regressions with Cray CCE when using the OpenACC \texttt{!\$acc routine seq} offloading directives.
In some isolated cases, the CCE compiler chooses not to inline the subroutine unless the \texttt{!\$acc routine seq} directive is replaced with the compiler hint \texttt{!\$DIR INLINEALWAYS <routine\_name>}.
The instances in which the CCE \texttt{!\$DIR INLINEALWAYS} hint is needed in place of the standard offload directive are not always obvious to developers, so automated benchmarking is a useful tool for identifying them.

Additional performance bottlenecks can be introduced in specific kernels where OpenACC struggles to produce efficient code.
Such bottlenecks can be due to large register usage or compiler choices.
For example, thread-private arrays that lack a known size at compile time require expensive memory reallocation for each independent loop when using the CCE compiler on AMD GPU devices.
Reordering the contents of large kernels can also result in performance regressions if the resulting code uses registers less efficiently for thread-private variables.
Due to its relatively small cache sizes, kernel ordering for efficient register use is especially relevant with AMD hardware.
These kernel-specific performance bottlenecks can be challenging to identify when developing code.
Automated benchmarking quickly identifies problems and provides a starting point for developers to determine the cause.

These performance regressions and bottlenecks identified by benchmarking demonstrate the value of automated benchmarks.
Desirable yet straightforward changes, such as refactoring, can cause substantial performance regressions that are not apparent during development.
MFC's automated benchmark suite offers a user-friendly approach to evaluating code performance and addressing identified regressions.

\begin{table*}
        \caption{
        Observed grindtime (called ``Time'') performance (nanoseconds per grid cell, equation, and right-hand side evaluation) for a standardized compressible CFD test problem.
        Results are shown for various CPU, GPU, and APU architectures.
        The best performing compiler is shown in each case, with GNU, Intel, NVIDIA, AMD, and CCE tested as appropriate.
        Smaller numbers are better.
        All results are collected using the compiler that performs best for each hardware.
        CPU results are parallelized via MPI, with each rank bound to a single core.
        The results can be interpreted as providing relative performance between single GPUs and single CPU sockets.
    }
    \label{tab:devicePerf}
    \begin{tabular}{lllll | lllll}
        Hardware & Type & Usage & Time &  & Hardware & Type & Usage & Time &   \\  \midrule
            NVIDIA GH200 & APU & 1 GPU & 0.32 & & NVIDIA A10 & GPU & 1 GPU & 4.3 \\ 
            NVIDIA H100 SXM5 & GPU & 1 GPU & 0.38 & & AMD EPYC 7713 & CPU & 64 cores & 5.0 \\ 
            NVIDIA H100 PCIe & GPU & 1 GPU & 0.45 & & Intel Xeon 8480CL & CPU & 56 cores & 5.0 \\ 
            AMD MI250X & GPU & 1 GPU & 0.55 & & Intel Xeon 6454S & CPU & 32 cores & 5.6 \\
            AMD MI300A & APU & 1 APU & 0.57 & &  Intel Xeon 8462Y+ & CPU & 32 cores & 6.2 \\
            NVIDIA A100 & GPU & 1 GPU & 0.62 & & Intel Xeon 6548Y+ & CPU & 32 cores & 6.6 \\ 
            NVIDIA V100 & GPU & 1 GPU & 0.99 & & Intel Xeon 8352Y & CPU & 32 cores & 6.6 \\ 
            NVIDIA A30 & GPU & 1 GPU & 1.1 & & Ampere Altra Q80-28 & CPU & 80 cores & 6.8 \\ 
            AMD EPYC 9965 & CPU & 192 cores & 1.2 & & AMD EPYC 7513 & CPU & 32 cores & 7.4 \\ 
            AMD MI100 & GPU & 1 GPU & 1.4 & & Intel Xeon 8268 & CPU & 24 cores & 7.5 \\ 
            AMD EPYC 9755 & CPU & 128 cores & 1.4 & & AMD EPYC 7452 & CPU & 32 cores & 8.4 \\ 
            Intel Xeon 6980P & CPU & 128 cores & 1.4 & & NVIDIA T4 & GPU & 1 GPU & 8.8 \\ 
            NVIDIA L40S & GPU & 1 GPU & 1.7 & & Intel Xeon 8160 & CPU & 24 cores & 8.9 \\ 
            AMD EPYC 9654 & CPU & 96 cores & 1.7 & & IBM Power10 & CPU & 24 cores & 10 \\ 
            Intel Xeon 6960P & CPU & 72 cores & 1.7 & & AMD EPYC 7401 & CPU & 24 cores & 10 \\ 
            NVIDIA P100 & GPU & 1 GPU & 2.4 & & Intel Xeon 6226 & CPU & 12 cores & 17 \\ 
            Intel Xeon 8592+ & CPU & 64 cores & 2.6 & & Apple M1 Max & CPU & 10 cores & 20 \\ 
            Intel Xeon 6900E & CPU & 192 cores & 2.6 & & IBM Power9 & CPU & 20 cores & 21 \\ 
            AMD EPYC 9534 & CPU & 64 cores & 2.7 & & Cavium ThunderX2 & CPU & 32 cores & 21 \\ 
            NVIDIA A40 & GPU & 1 GPU & 3.3 & & Arm Cortex-A78AE & CPU & 16 cores & 25 \\ 
            Intel Xeon Max 9468 & CPU & 48 cores & 3.5 & & Intel Xeon E5-2650V4 & CPU & 12 cores & 27 \\ 
            NVIDIA Grace CPU & CPU & 72 cores & 3.7 & & Apple M2 & CPU & \phantom{1}8 cores & 32 \\ 
            NVIDIA RTX6000 & GPU & 1 GPU & 3.9 & & Intel Xeon E7-4850V3 & CPU & 14 cores & 34 \\ 
            AMD EPYC 7763 & CPU & 64 cores & 4.1 & & Fujitsu A64FX & CPU & 48 cores & 63 \\ 
            Intel Xeon 6740E & CPU & 92 cores & 4.2 & & ~ & ~ & ~ & ~
    \end{tabular}
\end{table*}

\section{MFC as a tool for deployment and testing} \label{sec:exampleUses}

MFC's portability and well-defined performance metrics make it useful for evaluating emerging supercomputers and hardware-software combinations.
In the following sections, we present a standardized benchmark case with documented performance on nearly 50 different hardware platforms and strong and weak scaling results that serve as a reference for evaluating supercomputers.

\subsection{A standardized benchmark case} \label{sec:standardBenchmark}

Benchmark results are collected for a standardized three-dimensional (3D) CFD test problem.
The test problem simulates a two-phase flow (such as gas and liquid interaction) using a well-established mathematical model entailing a system of eight coupled PDEs.
The equations are solved using high-order numerical methods: fifth-order accurate WENO (weighted essentially non-oscillatory) spatial reconstructions for shock wave treatment, the HLLC (Harten--Lax--van~Leer contact) Riemann solver for finite volume flux computation, and a third-order accurate Runge--Kutta method for time advancement.
This combination of numerical methods is widely adopted in the CFD community for solving compressible and multiphase problems~\cite{devanna2023uranos1,bernardini2021streams,antonis2022ucns3d,bezgin2023jax, schmidmayer2020ecogen}.
The benchmark case is maintained in MFC's version control under \texttt{examples/3D\_performance\_test/} and can be executed on any target system to evaluate performance.

\Cref{tab:devicePerf} lists the grindtime performance of the standardized benchmark case run in double precision on a range of CPU, GPU, and APU architectures.
We observe similar grindtimes when solving related problems, such as the inviscid Euler equations (4~PDEs) and the six-equation multiphase flow model~\cite{saurel2009} (10~PDEs).
We report grindtimes using the compiler that produced the best results for each system.
Current tested compilers include Cray's CCE, NVIDIA's NVHPC, AMD's AOCC, Intel's OneAPI, and GNU GCC.
We benchmark nominally single-precision GPUs in double precision using hardware or compiler conversion.
Results for GPUs with more than one compute die (for example, the AMD MI250X and MI300A) are presented for the entire device.
Parallelism is achieved using MPI; for CPUs, each MPI rank is bound to a core.
CPUs may have more cores than the results reported for.
However, results are reported for the core count providing optimal performance.
Additional details on the hardware, compiler, and systems used are available in the MFC documentation\footnote{\url{mflowcode.github.io/documentation/md_expectedPerformance.html}}.
This collection of results is a reference for users to compare the performance of their hardware--software combinations against a range of systems spanning several generations.

\subsection{Weak scaling}

\begin{figure}
    \centering
    \Description{Weak scaling performance plot showing near-ideal scaling across five supercomputers}
    \includegraphics[]{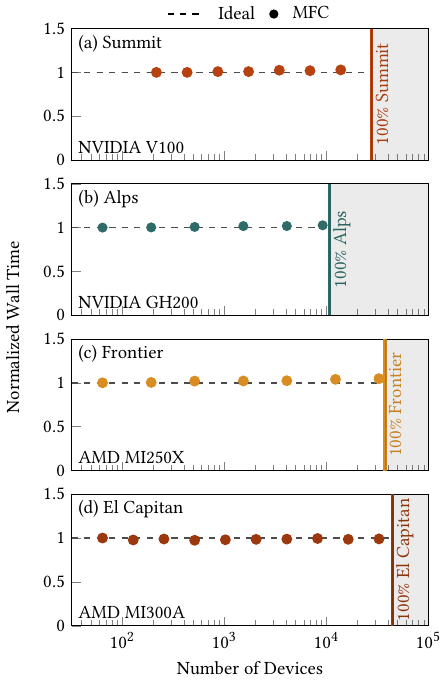}
    \caption{
        Weak scaling results for MFC on five flagship supercomputers.
        Near-ideal scaling is observed for multiple generations of AMD and NVIDIA hardware.
        \Cref{tab:wsNumbs} shows the details of each system's base case, limit case, and efficiency.
    }
    \label{fig:weakScaling}
\end{figure}

Weak scaling tests in MFC are performed using the same standardized case described in \cref{sec:standardBenchmark} with modified discretization and domain boundaries.
The domain boundaries and associated discretization are selected so that each MPI rank holds a local domain with a perfect cube of grid cells.
Uniform local discretization ensures uniform communication costs.
\Cref{tab:wsProbSize} lists example MPI discretizations and problem sizes used to collect the weak scaling results for OLCF Frontier in \cref{fig:weakScaling}.
The Frontier weak scaling test is conducted for a problem size of $200^3$ grid cells per MI250X GCD, which amounts to about \SI{16}{\giga\byte} of HBM2e memory.
The size is selected to be large enough to saturate the memory bandwidth of the MI250X GCDs to reach MFC's peak performance.
Weak scaling performance is measured using the previously defined grindtime metric.
The grindtime multiplied by the number of ranks should remain constant across problem sizes for ideal weak scaling, which we observe in all cases.
The file-per-process I/O strategy described in~\cite{wilfongSC24} reduces I/O overhead in weak scaling tests.
This approach is used when the number of MPI ranks exceeds $10^4$ or the total problem size exceeds $100$ billion spatially discretized grid cells.

\begin{table}
    \centering
    \caption{
        MPI decomposition and discretization details for a weak scaling test on OLCF Frontier.
        Each MPI rank has a local domain of $\bm{200 \times 200 \times 200}$ grid cells so that all halo exchanges are equivalent.
        Approximately \textbf{\SI{16}{\giga\byte}} of HBM2e memory is used per MI250X GCD, or one quarter of the available HBM2e memory.
    }
    \label{tab:wsProbSize}
    {\setlength{\tabcolsep}{4pt}
    \begin{tabular}{r c c S[table-format=3.2]}
        \toprule
        {\# Ranks} & {Decomposition} & {Discretization} & {\# Cells [B]} \\ 
        \midrule
        128   & $4 \times 4 \times 8$    & $800 \times 800 \times 1600$    & 1.02 \\
        384   & $6 \times 8 \times 8$    & $1200 \times 1600 \times 1600$  & 3.07 \\
        1024  & $8 \times 8 \times 16$   & $1600 \times 1600 \times 3200$  & 8.19 \\
        3072  & $12 \times 16 \times 16$ & $2400 \times 3200 \times 3200$  & 24.6 \\
        8192  & $16 \times 16 \times 32$ & $3200 \times 3200 \times 6400$  & 65.5 \\
        24576 & $24 \times 32 \times 32$ & $4800 \times 6400 \times 6400$  & 197 \\
        65536 & $32 \times 32 \times 64$ & $6400 \times 6400 \times 12800$ & 524 \\ \bottomrule
    \end{tabular}
    }
\end{table}

\Cref{fig:weakScaling} shows weak scaling results for MFC on four leadership-class supercomputers.
Three of the systems shown (OLCF~Summit~\cite{top500-summit}, OLCF~Frontier~\cite{top500-frontier}, and LLNL~El~Capitan~\cite{top500-elcap}) have held the number one position on the TOP500 list, and CSCS~Alps~\cite{top500-alps} is in the top ten and is the largest Grace~Hopper-based machine online at the time of writing.
We observe weak scaling efficiencies above 95\% for all systems, spanning three orders of magnitude in problem size and scaling to full systems.
\Cref{tab:wsNumbs} shows each system's base case, limit case, and efficiency.
MFC's consistent scaling performance across these systems makes it a suitable tool for testing weak scaling performance.

\begin{table}
    \centering
    \caption{
        Weak scaling efficiencies and device counts for the systems shown in \cref{fig:weakScaling}.
        Near-ideal efficiency is observed over three orders of magnitude in problem size for all systems across multiple generations of compute devices from NVIDIA and AMD.
    }
    \label{tab:wsNumbs}
    \begin{tabular}{r  c c c }
        \toprule
        System & Base case & Limit case  & Efficiency \\ 
        \midrule
        OLCF Summit & 216 GPUs & 13825 GPUs & 97\% \\
        CSCS Alps & \phantom{0}64 GPUs & \phantom{0}9200 GPUs & 97\% \\
        OLCF Frontier & 128 GCDs & 65536 GCDs & 95\% \\
        LLNL El Capitan & \phantom{0}64 GPUs & 32768 GPUs & 99\% \\ 
        \bottomrule
    \end{tabular}
\end{table}

\subsection{Strong scaling}

\begin{figure}
    \centering
    \Description{Strong scaling performance plots for OLCF Frontier and CSCS Alps showing speedup vs processor count.}
    \includegraphics[]{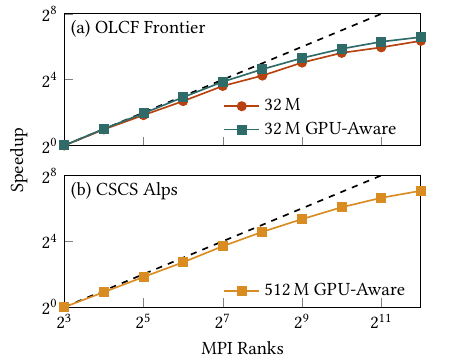}
    \caption{Strong scaling performance on (a) OLCF Frontier and (b) CSCS Alps.
    The speedup is calculated as the ratio of the grindtime for a given processor count to the grindtime of the 8~rank baseline.
    The impact of using GPU-Aware MPI to reduce communication overhead and improve strong scaling efficiency is shown in the OLCF Frontier results.
    Extension of near-ideal strong scaling behavior follows from using a larger base case on CSCS~Alps.
    }
    \label{fig:strongScaling}
\end{figure}

Scaling tests in MFC are performed using the standardized case of \cref{sec:standardBenchmark} with different spatial grid discretizations.
Performance is measured using the previously defined grindtime, which should scale inversely with the number of processors used for a constant problem size.
The problem size for the base case is selected to saturate the available GPU memory of 8~MPI ranks.
Saturating the device memory minimizes relative communication cost in the base case.
Using 8~MPI~ranks enables uniform MPI communication across the three spatial directions.
The maximum problem size per GCD on OLCF~Frontier is approximately \SI{32}{\million} grid cells.
So, an initial problem size of $634 \times 634 \times 634$ is used for scaling tests.
This results in \SI{31.9}{\million} grid cells per device in the 8~rank base case.
MFC supports GPU-Aware MPI (via RDMA), which is not enabled by default.
It can be enabled by adding \texttt{`rdma\_mpi': `T'} to the input case file when the machine supports it.
\Cref{fig:strongScaling}~(a) shows that GPU-aware MPI improves strong scaling efficiency on Frontier.

The strong scaling results for CSCS Alps in \cref{fig:strongScaling}~(b) use the alternative numerics described in~\cite{wilfong253} that enable use of a larger base case.
The larger base case is discretized using a $1600^3$ spatial grid, which results in \SI{512}{\million} grid cells per device in the 8~rank base case.
The larger base case shows preferable scaling efficiency on CSCS~Alps.
However, the trend is similar to that observed on OLCF~Frontier.
MFC's predictable scaling performance trends make it a valuable tool for evaluating the network performance of new supercomputers.

\section{Limitations of current work} \label{sec:limitations}

MFC's suitability as a tool for testing and benchmarking supercomputers has limitations.
The primary limitation is the reliance on compiler support for directives, either OpenMP or OpenACC, for offloading to non-CPU devices.
Currently, OpenMP and OpenACC provide limited or no support for more esoteric computing devices, such as Cerebras wafer-scale engines, Graphcore~IPUs, NextSilicon~Maverick-2, or quantum computers.
Researchers are starting to use these devices for stencil computations similar to those in MFC~\cite{brown2022cerebras, louw2021, song25}, though the current implementations rely on specialized compilers and lack portability.
Until these architectures are widely adopted and garner mainstream compiler support, MFC will be limited to testing CPU, GPU, and APU devices from major vendors.

\section{Conclusion} \label{sec:conclusion}

User-friendly applications are essential for evaluating the real-world performance of emerging supercomputers.
MFC's portability, well-defined performance metrics, and automated testing suite make it suitable for such evaluations.
Any user can run automated testing and benchmarking after a straightforward setup process.
This process abstracts away the details of the hardware and system.
The automated tools of MFC have been used to identify and report over 15 compiler and system bugs.
The results have identified numerous performance regressions when systems are updated.
MFC has also shown high-quality scaling performance across multiple generations of GPU hardware from NVIDIA and AMD.
These data serve as a reference for evaluating the performance of new computers.
These features and results establish MFC as a useful tool that nearly any user can use to test and benchmark supercomputers or incorporate into existing benchmarking suites.

\begin{acks}

SHB acknowledges support from the U.S. Department of Defense, Office of Naval Research under grant numbers N00014-22-1-2519 and N00014-24-1-2094, the Army Research Office under grant number W911NF-23-10324, the Department of Energy under DOE~DE-NA0003525 (Sandia~National~Labs, subcontract), the Oak Ridge Associated Universities (ORAU) Ralph E.\ Powe Junior Faculty Enhancement Award, and hardware gifts from NVIDIA and AMD.
Some computations were also performed on the Tioga, Tuolumne, and El~Capitan computers at Lawrence Livermore National Laboratory's Livermore Computing facility.
This research also used resources of the Oak Ridge Leadership Computing Facility at the Oak Ridge National Laboratory, which is supported by the Office of Science of the U.S.\ Department of Energy under Contract No.~DE-AC05-00OR22725 (allocation CFD154, PI Bryngelson).
This work used Delta and DeltaAI at the National Center for Supercomputing Applications and Bridges2 at the Pittsburgh Supercomputing Center through allocations PHY210084 and  PHY240200 (PI~Bryngelson) from the Advanced Cyberinfrastructure Coordination Ecosystem: Services \& Support (ACCESS) program, which is supported by National Science Foundation grants \#2138259, \#2138286, \#2138307, \#2137603, and \#2138296.
\end{acks}

\section*{Code Availability}

MFC is an open-source project; it is available under the MIT license.
The MFC source code is available at \url{https://github.com/MFlowCode/MFC}.
Additional information, documentation, and example simulations are available at \url{https://mflowcode.github.io}.

\bibliographystyle{ACM-Reference-Format}
\bibliography{main.bib}

\end{document}